# SIFT – File Fragment Classification Without Metadata

Shahid Alam
*Department of Computer Engineering*
*Adana Alparslan Turkes Science and Technology University*
Adana, Turkey
salam@atu.edu.tr

*Abstract*—A vital issue of file carving in digital forensics is type classification of file fragments when the filesystem metadata is missing. Over the past decades, there have been several efforts for developing methods to classify file fragments. In this research, a novel sifting approach, named SIFT (Sifting File Types), is proposed. SIFT outperforms the other state-of-the-art techniques by at least 8%. (1) One of the significant differences between SIFT and others is that SIFT uses a single byte as a separate feature, i.e., a total of 256 (0×00 – 0×FF) features. We also call this a lossless feature (information) extraction, i.e., there is no loss of information. (2) The other significant difference is the technique used to estimate inter-Classes and intra-Classes information gain of a feature. Unlike others, SIFT adapts TF-IDF for this purpose, and computes and assigns weight to each byte (feature) in a fragment (sample). With these significant differences and approaches, SIFT produces promising (better) results compared to other works.

## I. INTRODUCTION AND MOTIVATION

A filesystem contains metadata that expresses the real filesystem. The filesystem will also keep the physical locations on the storage device where each file is saved. In general, a filesystem allocates the several first sectors (disc blocks) of a storage device to retain the metadata such as the overall storage space, attributes of files, and their organization. The remaining sectors store the real content of the files. A sector is the smallest physical storage unit on a storage device with a typical size of 512 or 4096 bytes. Thus, a file might be dispersed in fragments at different physical addresses.

The metadata of a filesystem may not be available because of different damages or format operations to the storage device. In digital forensics, file carving is a number of steps taken and efforts to recover files on a storage device in part or whole without the metadata [1]. This is achieved by analyzing and classifying the raw data of file fragments located at sectors. Therefore, the identification of file fragment types is an essential problem in file carving. File fragment classification is also useful in other domains. Such as, in the network security realm of malware detection, type examination of packets' payload could advance the very early identification of malicious executable codes. Consequently, It is crucial to develop automated methods for file fragment classification. After the file fragment types are identified, the next step en-gages in ordering and merging the file fragments to reassemble the original file(s).

Most of the previous research on sifting file types can be categorized into (1) signature, (2) statistical, (3) artificial intelligence (AI), or (4) hybrid-based approaches [2]. A signature is an individualistic, unique, evidentiary attribute related to a file type. In signature-based approaches, comparison of known to unknown file fragment techniques are leveraged. The attributes of the file contents are utilized with statistical techniques in the second class of approaches. AI-based approaches generally leverage computational intelligence such as machine and deep learning methods. Applying a combination of these three approaches is considered a hybrid.

In this work, we propose a novel AI-based file segment classification method on a popular dataset [3]. At first, we preprocess the files in the dataset to separate the file fragments and their basic raw features. Afterward, Term Frequency and Inverse Document Frequency (TF-IDF) [4] technique is applied to gather the most decisive features among basic raw features. In other words, each raw feature is assigned a weight based on its TF-IDF, and the features having positive weights are selected. Finally, these weighted features are used to train and test a classifier to categorize the file fragments into file types. The results show that this approach is attainable and can achieve better outcomes. The major differences between the method proposed in this paper and previous other AI-based works [5]–[8] are (1) Lossless feature extraction. (2) Adaption of TF-IDF to estimate inter-Classes and intra-Classes information gain of a feature. Due to these new updates, our approach produces promising results compared to other works.

## II. SIFT – SYSTEM OVERVIEW

We first preprocess the files in a dataset to extract the fragments and their raw features. These features are then sifted through (examined thoroughly to isolate the most important) to mine and select important features. Each feature is assigned a weight according to its importance. These weighted features are then used by a classifier to classify the file types of the fragments. Figure 1 presents a high-level component overview of the proposed system SIFT. We further explain each of these components in detail in the following sections.

### A. Preprocessing and Feature Extraction

SIFT reads each file from a list of files and then preprocesses them. During preprocessing, SIFT: excludes files whose size < 2 fragment size; and keeps only one if more than one files have the same file name. After preprocessing SIFT extracts fragments at the byte level from each file. Each

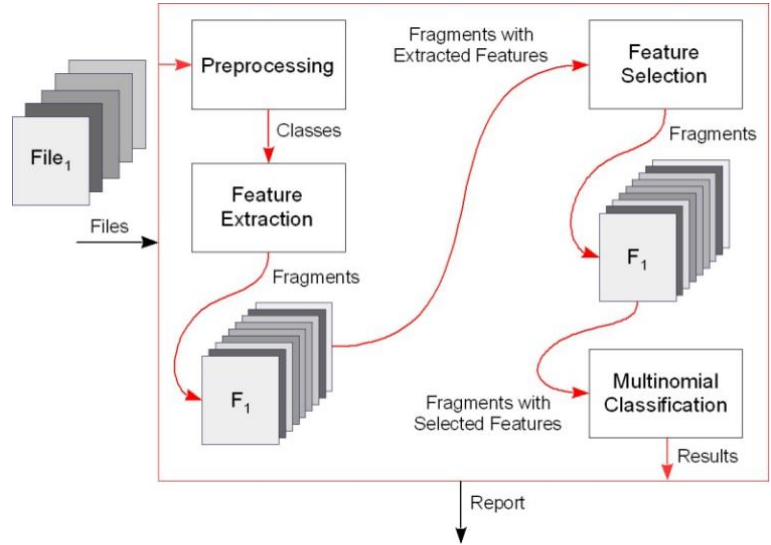

Fig. 1. Overview of the proposed system SIFT.

extracted fragment is of the same size R, that can vary from 25 – 212 = {32, 64, 128, 256, 512, 1024, 2048, 4096}. Catering a smaller fragment size than 512 bytes is because of resource constrained devices such as embedded systems and IoTs. At the byte level, we only have 256 different values. Therefore, the byte value ranges from 0×00 – 0×FF, and a total of R number of raw features are extracted for each fragment in a file. SIFT excludes the first fragment of a file, because of the presence of the header information that identifies the file type. Not all the files contain the equal size of fragments. To make each fragment the same size, the last partial fragment is filled with bytes from a randomly chosen fragment of the file as shown in Figure 2. Algorithms 1 and 2 list the steps to describe the process of extracting raw fragments from a list of files (dataset). Equation 1 formally define this set of raw fragments. We define a fragment as f = {b1, b2, b3, ..., bR}, where R is the size of the fragment; and b1 is byte number 1, b2 is byte number 2, and so on. A file in a dataset (set of files) is defined as file = {f1, f2, f3, ..., fm}, where m ∈ M; and f1 is fragment number 1, f2 is fragment number 2, and so on. Let M = {m1, m2, m3, ..., mN}: where N = number of files in a dataset; and m1 is the number of fragments extracted from file 1, m2 is the number of fragments extracted from file 2, and so on. Then, we define the set of extracted fragments F from a dataset as follows.

$$F = \bigcup_{i=1}^{N} \bigcup_{j=1}^{m \in M} \{f_{ij}, i\} \quad (1)$$

### B. Feature Selection

A fragment of size R consists of R number of bytes whose value ranges from 0×00 – 0×FF. Therefore, for a fragment, we select a total of 256 features and assign weight to each of them according to its importance in the fragment. Term Frequency and Inverse Document Frequency (TF-IDF) [4] is an experimental method in machine learning and data mining to segregate relevant features in a set of data. In our case, the set of data is the set of raw features defined by equation 1. TF-IDF computes the information gain of a term (in our case a) weighted by its occurrence of probability. We explain in the following, how we customize the TF-IDF weighting method and allocate weight to a byte (feature) and how this weight is applied to determine features from a fragment. We define TF-IDF of a byte $b_i \in f$ as follows:

$$TF_i = \frac{f_i}{n} \quad and \quad IDF_i = log\left(\frac{N}{K_i}\right)$$

where, $f_i$ is the number of times (recurrence) $b_i$ emerge in a fragment f; and $K_i$ is the number of all the fragments with $b_i$ in it. These definitions let us assign weight to a byte $b_i$ as follows:

$$W_i = TF_i \times IDF_i \quad (2)$$

### C. Multinomial Classification

The weight $W_i$ ranges from 0 – 1. We only keep $b_i$ if $W_i > 0$. For example, the byte 0×FB occurs several times in many fragments of type (Class) EPS, in the dataset used in this paper. There are a total of 20 Classes in the dataset used in this paper and are listed in Table I. The byte 0×FB gets a score > 0.98 for Class EPS and mostly 0 or < 0.25 for rest of the Classes. Similarly, the byte 0×30 gets a score > 0.98 for the Classes EPS, PS, and PDF, and mostly 0 or < 0.30 for the rest of the Classes. In this way, the weight assigning scheme, presented in equation 2, helps a classifier successfully classify a fragment by separating important/relevant features from the raw features, presented in equation 1.

## III. EMPIRICAL EVALUATION

We carried out an experimental evaluation to access the performance of SIFT and compare it with other state-of-the-art techniques. In this section, we present the dataset, metrics, experiments, results achieved, and discussion. We also compare SIFT with three other file fragment classification techniques. All the experiments were executed on Windows 8.1 running an Intel Core i-7 CPU @ 2.00 GHz with 8 GB of RAM.

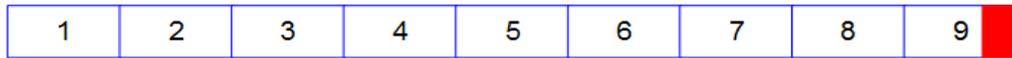

Fig. 2. Fragment extraction. As an example, there are 9 fragments of a file shown here. 8 are complete, whereas the last one is a partial fragment, filled to make it complete.

**Algorithm 1** Algorithm for extracting fragments from a list of files.

*Pre–conditions:* $files$ (List of files); $fragmentSize$ (Size of the fragment)
*Post–conditions:* Returns a dictionary of extracted $fragments$

1: **procedure** EXTRACTFRAGMENTS($files$, $fragmentSize$)
2:     $fragments$ = dict()    ; initialize the dictionary of fragments
3:     **for all** file $\in files$ **do**    ; iterate through the list of files
4:         $fragments$ = ExtractFromFile(file.contents(), $fragmentSize$)    ; extract fragments from the file
5:         $samples$[file.name] = fragments    ; store the extracted fragments under the filename
6:     **end for**
7:     **return** $fragments$    ; return the dictionary of fragments
8: **end procedure**

**Algorithm 2** Algorithm for extracting fragments from a file.

*Pre–conditions:* $fragmentSize$ (Size of the fragment); $fileContents$ (Contents of the file as bytes)
*Post–conditions:* Returns a list of extracted $fragments$

1: **procedure** EXTRACTFROMFILE($fileContents$, $fragmentSize$)
2:     $fileFragments$ = list()    ; initialize the list of fragments
3:     $c = n = fragmentSize$    ; $c$=start:$n$=end $\rightarrow$ start of the second fragment
4:     $numberOfFragments = \left\lfloor \frac{len(fileContents)}{fragmentSize} \right\rfloor$    ; number of fragments in the fileContents
5:     **for** $i = 1$ to $numberOfFragments$ **do**    ; extract these fragments from the fileContents
6:         $n = n + fragmentSize$    ; reset the end pointer
7:         $fragment = fileContents[c:n]$    ; extract the slice (fragment)
8:         $fileFragments$.append(fragment)    ; append the fragment to the list of fragments
9:         $c = n$    ; reset the start pointer
10:    **end for**
11:    $remainder = len(fileContents) \% fragmentSize$    ; calculate the size of any left over bytes
12:    **if** $remainder > 0$ **then**    ; extract the left over bytes
13:        $n = c + remainder$    ; $c$ is start and now $n$ is end of the left over bytes
14:        $i = random(0, numberOfFragments-1)$    ; randomly select a fragment
15:        $ac = i \times fragmentSize$    ; start of the selected fragment
16:        $an = addCurrent + fragmentSize - remainder$    ; end of the slice in the selected fragment
17:        $fragment = fileContents[c:n] + fileContents[ac:an]$    ; left over bytes + the slice
18:        $fileFragments$.append(fragment)    ; append the fragment to the list of fragments
19:    **end if**
20:    **return** $fileFragments$    ; return the list of fragments
21: **end procedure**

### A. Dataset

We randomly collected 20 file types from a publicly available dataset [3], which provides a more standardized dataset for digital forensics research. From these 20 file types, we extracted 47,482 samples (fragments). The file type distribution of these fragments is shown in Table I. To keep the evaluation unbiased we selected the same number (seven) of files from each Class (file type). Several previous researches have used either 512 [9]–[11] or 4096 [12]–[14] bytes as the fragment size. According to Penrose et al [13] 4096 bytes is a safe choice because all hard drive manufacturers have used this as their sector size since 2011. However, Axelsson et al. [9] observed that 512 bytes is a conservative choice. Therefore, we choose 512 bytes as the fragment size for our experiments.

### B. Evaluation Metrics

We define the evaluation metrics for our proposed model as follows. P is the number of fragments in the Class. N is the number of fragments in the other Classes. True positive (TP) is the number of fragments that are classified as positive. False positive (FP) is the number of fragments that are wrongly classified as positive. False negative (FN) is the number of fragments that are wrongly classified as negative. The TP rate (TPR) and FP rate (FPR) are defined as follows:

$$TPR = \frac{TP}{P} \quad \text{and} \quad FPR = \frac{FP}{N}$$

The Precision and F-Measure are defined as follows:

$$Precision = \frac{TP}{TP + FP}$$

$$F - Measure = \frac{2TP}{2TP + FP + FN}$$

### C. 10-fold Cross Validation

To evaluate the performance of SIFT, we used 10-fold cross validation. In this validation, we divided the dataset randomly into 10 equal size subsets. Out of these 10 subsets, 9 were used for training, and the remaining 1 was used for testing. This process is repeated 10 times. In each of these repetitions the 10 subsets were used exactly once for validation. This validation produces very systematized and precise testing results and also limits the problem of overfitting. Moreover, it gives an insight into how SIFT will generalize to an unknown dataset. We train the Random Forest classifier on the dataset and the results are shown in Table II. The confusion matrix is shown in Table III.

TABLE I FRAGMENT DISTRIBUTION OF THE 20 FILE TYPES (CLASSES). THERE ARE TOTAL 47,482 FRAGMENTS (SAMPLES) EACH OF SIZE 512 BYTES.

| Classes (File Types) | Number of Files | Number of Fragments |
|---|---|---|
| csv | 7 | 889 |
| dbase3 | 7 | 66 |
| doc | 7 | 2420 |
| eps | 7 | 5110 |
| gif | 7 | 701 |
| gz | 7 | 4470 |
| jpg | 7 | 924 |
| html | 7 | 613 |
| kmz | 7 | 1381 |
| log | 7 | 5346 |
| pdf | 7 | 2787 |
| png | 7 | 2704 |
| ppt | 7 | 3534 |
| ps | 7 | 3622 |
| swf | 7 | 1991 |
| text | 7 | 4671 |
| txt | 7 | 1374 |
| unk | 7 | 3264 |
| xls | 7 | 813 |
| xml | 7 | 802 |

### D. Discussion

The distribution of the results at the Class level are shown in Table V. Weight assigned using equation 2 to bytes in the fragments belonging to some of these Classes are shown in Table IV. To save space, only specific bytes are shown whose weights are much higher than other bytes. The average weight over all the fragments belonging to one Class is listed. For example, there are 889 fragments in the Class CSV (Table I), then there are 889 different weights assigned to the byte in each of these fragments.

Table IV lists the average of these 889 weights for the Class CSV and similarly for other Classes. These are the bytes that occur much more often in the fragments belonging to one of the specific Classes than any other Class. That means these are the attributes that will help successfully Classify these fragments. As shown in Table IV our weight assigning scheme successfully mines these and other attributes from the fragments (samples). This helps a classifier successfully classify a fragment of these Classes as shown in Table II. Our proposed approach is able to classify 15 out of the 20 Classes with a TPR > 70%. The results are poor for the Classes PDF, HTML, PPT, and SWF, because of the images embedded in these files. For example, we can see from the confusion matrices that 10.6% and 37.3% PDFs are wrongly classified as PNG and GZ, respectively.

TABLE II RESULTS OF THE 10-FOLD CROSS-VALIDATION.

| File Type | TPR | FPR | Precision | F-Measure |
|---|---|---|---|---|
| csv | 0.99 | 0.000 | 0.99 | 0.99 |
| dbase3 | 1.00 | 0.000 | 1.00 | 1.00 |
| doc | 0.75 | 0.008 | 0.84 | 0.79 |
| eps | 0.99 | 0.000 | 0.99 | 0.99 |
| gif | 0.35 | 0.000 | 0.98 | 0.51 |
| gz | 0.89 | 0.139 | 0.40 | 0.55 |
| jpg | 0.97 | 0.000 | 0.98 | 0.97 |
| html | 0.00 | 0.000 | 0.00 | 0.00 |
| kmz | 0.94 | 0.000 | 1.00 | 0.97 |
| log | 0.99 | 0.000 | 1.00 | 0.99 |
| pdf | 0.46 | 0.003 | 0.91 | 0.61 |
| png | 0.71 | 0.038 | 0.53 | 0.61 |
| ppt | 0.48 | 0.013 | 0.75 | 0.59 |
| ps | 0.99 | 0.001 | 0.98 | 0.99 |
| swf | 0.13 | 0.007 | 0.44 | 0.20 |
| text | 0.76 | 0.003 | 0.96 | 0.85 |
| txt | 0.92 | 0.000 | 0.99 | 0.96 |
| unk | 0.99 | 0.005 | 0.93 | 0.96 |
| xls | 0.81 | 0.001 | 0.94 | 0.87 |
| xml | 0.96 | 0.000 | 0.98 | 0.97 |
| Weighted Avg. | 0.80 | 0.018 | 0.84 | 0.79 |

Most of the fragments from PDF, PPT, SWF, and TEXT Classes are wrongly classified as GZ, which is a compressed file type. Because of compression, most of the bytes' weights, which makes fragments with similar bytes' weight from other Classes match with fragments from GZ Class.

(better) results. frequency in GZ is similar and are therefore assigned similar Related Work

TABLE III CONFUSION MATRIX OF THE 10-FOLD CROSS-VALIDATION.

| | csv | dbase3 | doc | eps | gif | gz | jpg | html | kmz | log | pdf | png | ppt | ps | swf | text | txt | unk | xls | xml |
|---|---|---|---|---|---|---|---|---|---|---|---|---|---|---|---|---|---|---|---|---|
| csv | 882 | 0 | 1 | 0 | 0 | 0 | 0 | 0 | 0 | 0 | 0 | 0 | 0 | 0 | 0 | 0 | 0 | 6 | 0 | 0 |
| dbase3 | 0 | 66 | 0 | 0 | 0 | 0 | 0 | 0 | 0 | 0 | 0 | 0 | 0 | 0 | 0 | 0 | 0 | 0 | 0 | 0 |
| doc | 0 | 0 | 1811 | 3 | 0 | 370 | 0 | 0 | 0 | 0 | 4 | 21 | 69 | 1 | 1 | 88 | 0 | 32 | 20 | 0 |
| eps | 0 | 0 | 0 | 5061 | 0 | 3 | 0 | 0 | 0 | 0 | 1 | 0 | 0 | 42 | 0 | 1 | 0 | 1 | 1 | 0 |
| gif | 0 | 0 | 1 | 0 | 245 | 192 | 0 | 0 | 0 | 0 | 13 | 153 | 63 | 0 | 34 | 0 | 0 | 0 | 0 | 0 |
| gz | 0 | 0 | 6 | 0 | 2 | 4008 | 0 | 0 | 0 | 0 | 18 | 287 | 70 | 0 | 54 | 2 | 0 | 23 | 0 | 0 |
| jpg | 0 | 0 | 2 | 0 | 0 | 0 | 899 | 0 | 0 | 1 | 0 | 0 | 0 | 0 | 0 | 1 | 3 | 8 | 0 | 10 |
| html | 0 | 0 | 1 | 0 | 0 | 452 | 0 | 0 | 0 | 0 | 5 | 111 | 15 | 0 | 25 | 1 | 0 | 3 | 0 | 0 |
| kmz | 0 | 0 | 0 | 0 | 0 | 40 | 0 | 0 | 1295 | 0 | 1 | 25 | 11 | 0 | 9 | 0 | 0 | 0 | 0 | 0 |
| log | 0 | 0 | 0 | 0 | 0 | 0 | 5 | 0 | 0 | 5338 | 0 | 0 | 0 | 2 | 0 | 0 | 0 | 0 | 0 | 1 |
| pdf | 0 | 0 | 3 | 4 | 0 | 1039 | 1 | 0 | 0 | 1 | 1274 | 295 | 82 | 3 | 76 | 4 | 0 | 4 | 0 | 1 |
| png | 0 | 0 | 1 | 0 | 1 | 675 | 0 | 0 | 0 | 0 | 14 | 1925 | 48 | 0 | 39 | 0 | 0 | 1 | 0 | 0 |
| ppt | 1 | 0 | 75 | 1 | 1 | 1291 | 0 | 0 | 0 | 0 | 27 | 328 | 1696 | 0 | 77 | 21 | 0 | 7 | 9 | 0 |
| ps | 0 | 0 | 0 | 5 | 0 | 0 | 0 | 0 | 0 | 0 | 0 | 0 | 0 | 3610 | 0 | 0 | 0 | 7 | 0 | 0 |
| swf | 0 | 0 | 6 | 0 | 1 | 1158 | 0 | 0 | 0 | 0 | 24 | 409 | 113 | 0 | 258 | 10 | 0 | 10 | 2 | 0 |
| text | 0 | 0 | 165 | 2 | 0 | 755 | 1 | 0 | 0 | 0 | 11 | 90 | 41 | 2 | 6 | 3555 | 0 | 36 | 7 | 0 |
| txt | 0 | 0 | 2 | 0 | 0 | 0 | 4 | 0 | 0 | 0 | 0 | 0 | 0 | 3 | 0 | 0 | 1271 | 94 | 0 | 0 |
| unk | 0 | 0 | 0 | 0 | 0 | 0 | 0 | 0 | 0 | 0 | 2 | 0 | 0 | 3 | 0 | 0 | 12 | 3247 | 0 | 0 |
| xls | 0 | 0 | 86 | 1 | 1 | 1 | 0 | 0 | 0 | 0 | 1 | 0 | 40 | 6 | 0 | 16 | 0 | 1 | 660 | 0 |
| xml | 0 | 0 | 5 | 0 | 0 | 0 | 10 | 0 | 0 | 0 | 0 | 0 | 0 | 1 | 0 | 2 | 0 | 10 | 0 | 774 |

TABLE IV WEIGHT ASSIGNED USING EQUATION 2 TO BYTES IN THE FRAGMENTS BELONGING TO THE 5 CLASSES. TO SAVE SPACE, ONLY SPECIFIC BYTES ARE SHOWN WHOSE WEIGHTS ARE MUCH HIGHER THAN OTHER BYTES.

| Class | Byte value in hex | Symbol | Description | Weight Assigned (Averaged) |
|---|---|---|---|---|
| CSV | 0×2C | , | Comma | 0.663 |
| CSV | 0×22 | " | Double quotes | 0.335 |
| DBASE3 | 0x20 | | Space | 0.939 |
| EPS | 0×48 | 0 | Zero | 0.462 |
| XML | 0×3C | < | Open angled bracket | 0.185 |
| XML | 0×3E | > | Close angled bracket | 0.181 |
| LOG | 0×3A | : | Colon | 0.160 |

### E. Comparison with Other Works

Table VI provides a comparison of SIFT with the other three state-of-the-art techniques. SIFT outperforms the others by 8% – 19%. (1) One of the major differences between SIFT and others is that SIFT uses a single byte as a separate feature, i.e., a total of 256 (0×00 – 0×FF) features. We also call this a lossless feature (information) extraction, i.e., there is no loss of information. (2) The other major difference is the technique used to estimate inter-Classes and intra-Classes information gain of a feature. Unlike others, SIFT adapts TF-IDF for this purpose and computes and assigns weight to each byte (feature) in a fragment (sample). With these major differences and approaches, SIFT produces promising

TABLE V DISTRIBUTION OF THE INDIVIDUAL AND COMBINED RESULTS AT THE CLASS LEVEL.

| TPR (%) | Classes |
|---|---|
| 91 – 100 | CSV, DBASE3, EPS, JPG, KMZ, LOG, TXT, UNK, XML |
| 71 – 90 | DOC, GZ, PNG, TEXT, XLS |
| 0 – 70 | GIF, HTML, PDF, PPT, SWF |

## IV. RELATED WORK

In this section, we briefly highlight recent research works on file fragment classification in the context of file carving. We divide these works into three popular categories.

TABLE VI COMPARISON OF THE MODEL SIFT PROPOSED IN THIS PAPER AND THREE OTHER STATE-OF-THE-ART MODELS.

| Model | TPR | Number of fragments (samples) | Size of the fragment | Number of Classes | Technique used |
|---|---|---|---|---|---|
| SIFT | 80% | 47,482 | 512 bytes | 20 | Adapted TF-IDF for assigning weights to each byte (feature) in a fragment (sample) |
| Haque et al. | 72% | 87,500 | 4096 bytes | 35 | Byte2Vec embeddings – extension of Word2Vec and Doc2Vec |
| Bhatt et al. | 67% | 14,000 | 512 bytes | 14 | Ten features, such as entropy, bigram distribution, hamming weight, and mean byte value, etc. |
| Wang et al. | 61% | 270,000 | 512 bytes | 18 | Continuous sequence (n–gram) of bytes of different sizes |

*A. Signature Based Approaches*

Signature-based approaches exploit the possible embedded signatures [2] in file headers and footers. Roussev et al. [15] suggested using sdhash real-time digital forensics and triage. Breitinger et al. [16] proposed essentially the use of similarity preserving hashing (SPH). Afterward, Lillis et al. [17] increased the speed of lookup of signatures through hierarchical Bloom filter tress.

Earlier, Garfinkel et al. [18] and Dandass et al. [19] used hash-values of fragments to identify individual files with the same fragments. Some changes on MD5, SHA1 to CRC32 hashing algorithms were also favored to compute hash values. Furthermore, Garfinkel et al. [20] explored a speedier way to compare master files with image files by using maps.

*B. Statistical Approaches*

Conti et al. [21], [22] extract statistical features, such as ChiSquare, Shannon entropy, Arithmetic mean, and Hamming weight to analyze the low-level binary data. 1000 fragments, where fragment size is 1 KB, in each category are analyzed. Statistical features are determined according to the spread of data fragments by original fragment class. The authors observed that the high entropy and machine code primitive types are more compactly clustered than the bitmap samples.

Calhoun et al. [23] extract statistical features, such as entropy and frequency of ASCII codes to classify graphic files, JPG, GIF, etc. They achieve promising results (83% accuracy) but are only applicable to graphic types. Veenman et al. [24] extract statistical features, such as histogram, and entropy to classify disc images. multi-class and two-class perception experiments were run using a dataset of 450 MB, achieving an overall accuracy of 0.45 which is quite modest. While ZIP files were classified with only 18% accuracy, HTML and JPEG files resulted in 98% accuracy.

Karresand et al. [25] proposed Oscar that computes the difference of the ASCII values between two consecutive bytes as a rate of change to classify file types. Oscar only performs well on JPG file types. Centroids, representing the mean and standard deviation of the byte frequency distribution of distinct file types, are the base for the Oscar method. The distance between the centroid and sample data fragments is assigned with a weighted quadratic distance metric. Upon the distance drop below a threshold, the sample is classified as possibly belonging to the modeled file type. Li et al. [26] use a 1gram binary distribution for file fragment classification on files collected from the Internet using a general search of a file type on Google. Promising results are obtained by using a one-centroid and multi-centroid file type model. McDaniel et al. [27] developed a file fingerprint for file type detection. They use the byte frequency analysis, byte frequency correlation, and file header/trailer information to generate the file fingerprint.

*C. Artificial Intelligence (AI) Based Approaches*

Haque et al. [5] proposed a model that extends Doc2Vec and Word2Vec embeddings to bytes and fragments, respectively. This model is referred to as Byte2Vec. Fragments, each of size 4096 bytes, are extracted from each file. These fragments are vectorized using the Byte2Vec embeddings. After extracting features using the Byte2Vec model, the k-nearest neighbor is used for classification. During testing the Byte2Vec model achieved an accuracy of 72% and a TPR of 72%.

Bhatt et al. [6] proposed a hierarchical machine-learning based model for the classification of file fragments. They use SVM as a base classifier. A total of ten features, such as entropy and bigram distribution, hamming weight, mean byte value, etc., are extracted from each fragment. Their model achieved an accuracy of 67.78% and a TPR of 67%. Chen et al. [7] proposed a method that first converts a fragment to a grayscale image for extracting high dimensional features, and then builds a convolution neural network model for the classification of fragments. During experiments, the model achieved an accuracy of 70.9%.

Wang et al. [8] use sparse coding for automatic feature extraction. Sparse coding extracts features based on how well these can be used to reconstruct the original data. For this purpose, they use a continuous sequence of bytes (n-grams) of different sizes. Linear SVM was used for classification. Their approach achieved an accuracy of 61.31% and a TPR of 60.99%.

Most of the previous works classify only special types of fragments, such as graphic types (JPG, GIF, PNG, etc.). Some of the current methods do not perform well for fragments with high entropy, because they do not have discernible patterns to exploit. Our approach does not have such a limitation because of lossless feature extraction that makes it possible to successfully classify different fragment types except a few.

## V. Limitations and Future Work

Computer files are often embedded with other files, such as images, PNG and JPG, etc., embedded in PDF and PPT file types. Some fragments (image type) of these files will be classified as the other Class (image). In this case, SIFT is not able to correctly identify these fragment types.

The distributions of different files in the dataset GovDocs, used in this paper, is not the same. This phenomenon can cause a bias during classification. For example, if the number of files in a class is very small than others, then it can affect the accuracy of a classifier. In this paper, we do not optimize the dataset itself. This can lead to a bias that may affect the accuracy of SIFT. In the future work, we will focus on optimizing the datasets to further improve classification accuracy of SIFT.

## VI. Conclusion

An essential issue in file carving is the recognition of file fragment types. In this paper, we propose a file fragment type identification method, named SIFT, based on the TF-IDF technique to assign a weight for each byte (feature) to select important features in a fragment. We used 512-byte segments. Then, we investigated the multinomial classifier Random Forest to evaluate the performance on a popular and publicly available dataset by 10-fold cross-validation. SIFT achieves a TPR of 80% during 10-fold cross-validation. Compared to other state-of-the-art methods that uses the same dataset as used in this paper, SIFT outperforms by 8% – 19%.